\documentclass[12pt,reqno]{amsart}
\usepackage{graphics,amsmath,amssymb,epsfig,stmaryrd,color}
\usepackage{subfigure,amsfonts,geometry,array}
\usepackage{graphicx}
\usepackage{amsthm}
\usepackage{flafter}

\usepackage{hyperref}

\usepackage[
backend=biber,
style=apa,
sorting=nyt
]{biblatex}
\DeclareSourcemap{
  \maps[datatype=bibtex]{
    \map{
      \step[fieldset=issn, null]
      \step[fieldset=doi, null]
      \step[fieldset=url, null]
      \step[fieldset=urldate, null]
    }
  }
}

\hypersetup{
    colorlinks=true,
    urlcolor=blue,
    linkcolor=blue,
    citecolor=blue
}

\newtheorem{theorem}{Theorem}
\theoremstyle{plain}

\newtheorem{assumption}{Assumption}

\newtheorem{definition}{Definition}
\newtheorem{example}{Example}

\newtheorem{proposition}{Proposition}

\numberwithin{equation}{section}

\begin{document}
\title{Dilation Bootstrap}
\author{Alfred Galichon and Marc Henry}
\noindent\date{First version: May 2006. This version: October 22, 2012. Correspondence address: Alfred Galichon, \'Ecole polytechnique, D\'epartment d'\'economie, 91128 Palaiseau, France. Marc Henry, Universit\'{e} de Montr\'{e}al, D\'{e}partement de sciences \'{e}conomiques, 3150, Jean-Brillant, Montr\'eal, Qu\'ebec H3C 3J7, Canada}

\maketitle

\begin{abstract}
We propose a methodology for constructing confidence regions with
partially identified models of general form. The region is obtained
by inverting a test of internal consistency of the econometric
structure. We develop a dilation bootstrap methodology to deal with
sampling uncertainty without reference to the hypothesized economic
structure. It requires bootstrapping the quantile process
for univariate data and a novel generalization of the latter to higher
dimensions. Once the dilation is chosen to control the confidence
level, the unknown true distribution of the observed data can be
replaced by the known empirical distribution and confidence regions
can then be obtained as in \cite{GH:2008a} and \cite{BMM:2008}.
\end{abstract}

\vskip6pt \noindent {\scriptsize JEL Classification: C15, C31.
\\Keywords: Partial identification, dilation bootstrap, quantile process, optimal matching.}

\newpage
\section*{Introduction}

In several rapidly expanding areas of economic research, the
identification problem is steadily becoming more acute. In policy
and program evaluation \cite{Manski:90} and more general
contexts with censored or missing data \cite{SV:2005},
\cite{MM:2005} and measurement error \cite{CHT:2005},
ad hoc imputation rules lead to fragile inference. In demand
estimation based on revealed preference \cite{BBC:2005} the
data is generically insufficient for identification. In the analysis
of social interactions \cite{BD:2001},
\cite{Manski:2004}, complex strategies to reduce the large
dimensionality of the correlation structure are needed. In the
estimation of models with complex strategic interactions and
multiple equilibria \cite{BV:85}, \cite{Tamer:2003},
assumptions on equilibrium selection mechanisms may not be available
or acceptable.

More generally, in all areas of investigation with
structural data insufficiencies or incompletely specified economic
mechanisms, the hypothesized structure fails to identify a unique
possible data generating mechanism for the data that is actually
observed. In such cases, many traditional estimation and testing
techniques become inapplicable and a framework for inference in
incomplete models is developing, with an initial focus on estimation
of the set of structural parameters compatible with true data
distribution (hereafter {\em identified set}). A question of
particular relevance in applied work is how to construct valid
confidence regions for the identified set. Formal methodological
proposals abound since the seminal work of \cite{CHT:2007},
but computational efficiency is still a major concern.

In the
present work, we propose a methodology that clearly distinguishes
how to deal with sampling uncertainty on the one hand, and model
uncertainty on the other, so that unlike previous methodological
proposals, search in the parameter space is conducted only once,
thereby greatly reducing the computational burden. The key to this
separation is to deal with sampling variability without any
reference to the hypothesized structure, using a methodology we call
the {\em dilation method}. This consists in dilating each point in
the space of observable variables in such a way that the empirical
probability (which is known) of a dilated set dominates the true
probability (which is unknown) of the original set (before dilation).
The unknown
true probability (i.e. the true data generating mechanism) is then
removed from the analysis, and we can proceed as if the problem were
purely deterministic, hence apply the methods proposed in \cite{GH:2008a} and \cite{BMM:2008}.

To construct confidence regions of level $1-\alpha$ for the
identified set, such a dilation $y\rightrightarrows J(y)$ (where $\rightrightarrows$ denotes a one-to-many map) must
satisfy $\tilde{Y}^\ast\in J(\tilde{Y})$ a.s. for some pair of
random vectors $(\tilde{Y}^\ast,\tilde{Y})$, with probability
$1-\alpha$, where $\tilde{Y}$ is drawn from the true distribution of
observable variables and $\tilde{Y}^\ast$ is drawn from the
empirical distribution relative to the observed sample. We propose a
dilation bootstrap procedure to construct $J$, in which bootstrap
realizations $Y_j^b$, $j=1,\ldots,n$ are matched one-to-one with the
original sample points $Y_j$, $j=1,\ldots,n$ so as to minimize
$\eta_n^b=\max_{j=1,\ldots,n}\|Y_j^b-Y_{\sigma(j)}\|$, where the
permutation $\sigma$ defines the matching. The $\alpha$ quantile of
the distribution of $\eta_n^b$ then defines the radius of the
dilation.

When the observable $Y$ is a random variable, the dilation bootstrap
relies on bootstrapping the quantile process, as proposed by
\cite{DG:92}. However, bootstrapping the quantile process
relies on order statistics and had no higher dimensional
generalization to date. This is now provided by the dilation
bootstrap, which removes the constraint on dimension through the
appeal to optimal matching. Although the problem of finding minimum
cost matchings (called the {\em assignment} or {\em marriage
problem}) is very familiar to economists, as far as we know, its
application within an inference procedure is unprecedented.

The rest of the paper is organized as follows. The next section
describes the econometric framework and introduces the {\em
Composition Theorem} and the dilation method the latter justifies.
Section~\ref{subsection: confidence region} discusses the
application of the Composition Theorem to constructing confidence
regions for partially identified parameters.
Section~\ref{subsection: dilation bootstrap1} presents the bootstrap
feasible dilation and its theoretical underpinnings.
Section~\ref{section: simulation} presents simulation evidence on
the performance of the dilation bootstrap in comparison with
alternative methods. Section~\ref{section: extensions}
explains how the method extends to higher dimensions and discrete choice and the last section concludes.

\section{Dilation method and Composition Theorem}
\label{section: composition}

We consider the problem of inference on the structural parameters of
an economic model, when the latter are (possibly) only partially
identified. The economic structure is defined as in
\cite{Jovanovic:89}, which generalizes \cite{KR:50}.
Variables under consideration are divided into two groups. Latent
variables $U$ capture unobserved heterogeneity in the model. They
are typically not observed by the analyst, but some of their
components may be observed by the economic actors. Observable
variables $Y$ include outcome variables and other observable
heterogeneity. They are observed by the analyst and the economic
actors. We call \emph{observable distribution} $P$ the true
probability distribution generating the observable variables, and
denote by $\nu$ the probability distribution that generated the
latent variables $U$. The econometric structure under consideration
is given by a binary relation between observable and latent
variables, i.e. a subset of $\mathcal{Y}\times\mathcal{U}$, which can be
written without loss of generality as a correspondence from $\mathcal{U}$ to $\mathcal{Y}$.

\begin{assumption}[Econometric specification]\label{assumption: specification}
Observable variables $Y$, with realizations $y\in \mathcal{Y}\subseteq\mathbb{R}^{d_y}$ and latent variables $U$, with
realizations $u\in \mathcal{U}\subseteq\mathbb{R}^{d_u}$, are defined
on a common probability space $(\Omega,\mathcal{F},\mathbb{P})$ and
satisfy the relation: $Y\in G(U)\subseteq\mathcal{Y}$ almost
surely.\end{assumption}

\begin{example}[Revealed Preferences]
This approach is particularly well suited to revealed preference
analysis. Suppose $X$ is the vector of observed choices made by an
agent, possibly over several periods. Let $Z$ be a vector of
observable variables defining the environment in which the agent
made their choices. Call $Y=(X,Z)$ the vector of all observable
variables. Suppose the agent maximized a utility
$u(X,Z,U\vert\theta)$ under constraints $g(X,Z,U\vert\theta)\leq0$
(budget constraints, etc...), where $\theta$ is a vector of
structural parameters (including elasticities, risk aversion,
etc...) and $U$ a random vector describing unobserved heterogeneity.
Call $D(U,X\vert\theta)$ the demand correspondence, i.e. the set of
utility maximizing choices. Then we can define $G(U\vert\theta)$ by
$Y\in G(U\vert\theta)$ if and only if $X\in D(U,X\vert\theta)$, and
$G(U\vert\theta)$ exhausts all the information embodied in the
utility maximization model.\label{example: revealed
preferences}\end{example}

\begin{example}[Games]
Another family of examples of our framework arises with parametric
games. Let $N$ players with observable characteristics
$X=(X_1,\ldots,X_N)$ and unobservable characteristics
$U=(U_1,\ldots,U_N)$ have strategies $Z=(Z_1,\ldots,Z_N)$ and
payoffs parameterized by $X,U,Z$ and $\theta$. For a given choice of
equilibrium concept in pure strategies, call
$\mathcal{C}(X,U,\theta)$ the equilibrium correspondence, i.e. the
set of pure strategy equilibrium profiles. Then the empirical
content of the game is characterized by
$Z\in\mathcal{C}(X,U,\theta)$, which can be equivalently rewritten
$Y\in G(U;\theta)$ with $Y=(Z,X)$.\end{example}

We assume a parametric structure for the unobserved heterogeneity
and the model linking unobserved heterogeneity variables to
observable ones.

\begin{assumption}[Correspondence]\label{assumption: correspondence}
The correspondence $G:\mathcal U\rightrightarrows\mathcal Y$ is
known by the analyst up to a finite dimensional vector of parameters
$\theta\in\Theta\subseteq\mathbb{R}^{d_\theta}$. It is denoted
$G(\cdot;\theta)$. For all $\theta\in\Theta$, $G(\cdot;\theta)$ is
measurable (i.e. the set $\{u:\;G(u;\theta)\cap A\ne\varnothing\}$ is
measurable for each open subset $A$ of $\mathcal{Y}$) and has non
empty and closed values.\end{assumption}

Note that the measurability and closed values assumptions are very
mild conditions. The assumption that the correspondence is
non-empty, however, may be restrictive. In the revealed preferences
example, we require that the demand correspondence be non empty. In
the games example, we require existence of equilibrium.

\begin{assumption}[Latent variables] \label{assumption: parametric}
The distribution $\nu$ of the unobservable variables $U$ is assumed
to belong to a parametric family $\nu(\cdot\vert\theta)$,
$\theta\in\Theta$. The same notation is used for the parameters of
$\nu$ and $G$ to highlight the fact that they may have components in
common.\end{assumption}

The pair of random vectors $(Y,U)$ involved in the model is
generated by a probability distribution, that we denote $\pi$. Since
the vector $U$ is unobservable, the probability distribution $\pi$
is not directly identifiable from the data. However, the econometric
model imposes restrictions on $\pi$. The distribution of its
component $Y$ is the observable distribution $P$. The distribution
of its component $U$ is the hypothesized probability distribution
$\nu(\cdot\vert\theta)$. Finally, the joint distribution is further
restricted by the fact that it gives mass 0 to the event that the
relation $Y\in G(U\vert\theta)$ is violated. For any given value of
the structural parameter vector $\theta$, a joint distribution
satisfying all these restrictions may or may not exist. If it does,
it is generally non unique. The identified set $\Theta_I$ is the
collection of values of the structural parameter vector $\theta$ for
which such a joint probability distribution does indeed exist.

\begin{itemize}\item If $\Theta_I=\varnothing$, the model is rejected.
\item If $\Theta_I$ is a singleton, the parameter vector $\theta$ is point identified.
\item Otherwise, the parameter $\theta$ is set identified.\end{itemize}

The set $\Theta_I$, first formalized in this way in
\cite{GH:2006a} is sometimes called ``sharp identification
region'' to emphasize the fact that it exhausts all the information
on the parameter available in the model. No value
$\theta\in\Theta_I$ could be rejected on the basis of the knowledge
of the model and the observable distribution $P$ only. Take a
parameter value $\theta\in\Theta$. It belongs to the identified set
$\Theta_I$ if and only if there exists a joint distribution
satisfying the required restrictions, in other words, if and only if
there exists a ``version'' of $U$, i.e. a random vector $\tilde{U}$
with the same distribution as $U$, namely $\nu(\cdot\vert\theta)$,
such that $Y\in G(\tilde{U}\vert\theta)$ with probability 1. Hence,
denoting by $X\sim\mu$ the statement ``the random vector $X$ has
probability distribution $\mu$,'' we can characterize the identified
set in the following way, which we take as our formal definition.

\begin{definition}[Identified set]\label{definition: identified set}
\[\Theta_I=\left\{\theta\in\Theta\;\vert\;\;\exists\tilde{Y}\sim
P,\;\tilde{U}\sim\nu(\cdot\vert\theta):\;\mathbb{P}(\tilde{Y}\notin
G(\tilde{U}\vert\theta))=0 \right\}.\] \end{definition}

Our inference method on the identified set will be based on a
general way of combining sources of uncertainty (sampling
uncertainty or data incompleteness) by composition of
correspondences. Suppose the probability measure $Q$ on
$\mathcal{Y}$ is the known distribution of a random vector $Z$ and
that it is related to the true unknown distribution $P$ of the
observed variables $Y$ by the following relation:

\begin{assumption}[Dilation] \label{assumption: dilation}
There exists a correspondence
$J:\mathcal{Y}\rightrightarrows\mathcal{Y}$ such that
$\mathbb{P}(\tilde{Z}\notin J(\tilde{Y}))\leq\beta$ for some
$\tilde{Z}\sim Q$, $\tilde{Y}\sim P$ and
$0\leq\beta<1$.\end{assumption}

Assumption~\ref{assumption: dilation} characterizes the additional
level of indeterminacy the analyst faces. The structural model is
incomplete in the sense that the relation between unobserved
heterogeneity $U$ and outcomes $Y$ is a many-to-many mapping. In
addition, due to observability issues or sampling uncertainty, the
distribution of outcomes $P$ is unknown and the relation between
true outcome $Y$ and a variable $Z$ that we can simulate is also
many-to-many.

\begin{example}[Measurement error]\label{example: measurement} Suppose true outcome $Y$ is mismeasured
as $Z=Y+\epsilon$ and nothing is known about measurement error
$\epsilon$ except that it is small, i.e. $\|\epsilon\|\leq\eta$ for
some $\eta>0$, with a degree of confidence $1-\beta$. In that case,
Assumption~\ref{assumption: dilation} holds with the correspondence
$J$ defined by $J(y)=B(y,\eta)$ for all $y\in\mathcal{Y}$, where
$B(y,\eta)$ is the closed ball centered at $y$ with radius $\eta$.
\end{example}

\begin{example}[Censored outcomes]\label{example: censored} Suppose the true outcome $Y$ is
reported with censoring as $Z=J(Y)$, where $J(y)$ returns the
minimum of $y$ and an upper bound $B>0$.
Assumption~\ref{assumption: dilation} is satisfied with $\beta=0$.
\end{example}

The following theorem shows how the two levels of uncertainty can be
combined without loss of information.\footnote{The current proof of Theorem~\ref{theorem: composition}, suggested by Alexei Onatski, is shorter and simpler than our original proof in previous versions of the paper. We are responsible for any remaining errors.}

\begin{theorem}[Composition Theorem]\label{theorem: composition}
Under assumptions~\ref{assumption: specification}
to~\ref{assumption: dilation}, there exist $\tilde{Z}\sim Q$ and
$\tilde{U}\sim \nu$ such that $\mathbb{P}(\tilde{Z}\notin J\circ
G(\tilde{U}\vert\theta))\leq\beta$.\end{theorem}

Theorem~\ref{theorem: composition} implies that when the
distribution $P$ of outcomes is unknown, the infeasible identified
set $\Theta_I$ can be replaced by a feasible identified set
\[\tilde\Theta_I=\left\{\theta\in\Theta\;\vert\;\;\exists\tilde{Z}\sim
Q,\;\tilde{U}\sim\nu(\cdot\vert\theta):\;\mathbb{P}(\tilde{Z}\notin
J\circ G(\tilde{U}\vert\theta))\leq\beta \right\}.\]

\begin{proof}[Proof of Theorem~\ref{theorem: composition}]
Under Assumptions~\ref{assumption: specification}
and~\ref{assumption: parametric}, there is a pair $(Y,U)$ such that
$Y\sim P$ and $U\sim\nu(\cdot\vert\theta)$ and $Y\in
G(U\vert\theta)$ almost surely. Equivalently, the minimum over all
pairs $(\tilde{Y},\tilde{U})$, with $\tilde{Y}\sim P$ and
$\tilde{U}\sim\nu(\cdot\vert\theta)$, of the quantity
$\mathbb{E}(1\{\tilde{Y}\notin G(\tilde{U}\vert\theta)\})$ is zero.
By proposition~1 of \cite{GH:2006d} (hereafter denoted P1) , the
latter is equivalent to
\begin{eqnarray}\sup\left(
P(A)-\nu(\{u\in\mathcal{U}:\;G(u\vert\theta)\cap
A\ne\varnothing\}\vert\theta)\right)=0,\label{equation:
dual1}\end{eqnarray} where the sup is over all Borel subsets $A$ of
$\mathcal{Y}$. Similarly, by Assumption~\ref{assumption: dilation},
the minimum over all pairs $(\tilde{Z},\tilde{Y})$, with
$\tilde{Z}\sim Q$ and $\tilde{Y}\sim P$, of the quantity
$\mathbb{E}(1\{\tilde{Z}\notin J(\tilde{Y})\})$ is smaller than or
equal to $\beta$. By P1 the latter is equivalent to
\begin{eqnarray}\sup\left(
Q(A)-P(\{y\in\mathcal{Y}:\;J(y)\cap
A\ne\varnothing\}\vert\theta)\right)\leq\beta.\label{equation:
dual2}\end{eqnarray} Denote $J^{-1}(A)=\{y\in\mathcal{Y}:\;J(y)\cap
A\ne\varnothing\}$. By (\ref{equation: dual1}), we have
$P(J^{-1}(A))\leq\nu(\{u\in\mathcal{U}:\;G(u\vert\theta)\cap
J^{-1}(A)\ne\varnothing\}\vert\theta)$ for all Borel subsets $A$ of
$\mathcal{Y}$. Hence, (\ref{equation: dual2}) yields
\[\sup\left(Q(A)-\nu(\{u\in\mathcal{U}:\;G(u\vert\theta)\cap
J^{-1}(A)\ne\varnothing\}\vert\theta)\right)\leq\beta,\] Hence
\begin{eqnarray}\sup\left(Q(A)-\nu(\{u\in\mathcal{U}:\;J\circ
G(u\vert\theta)\cap A\ne\varnothing\}\vert\theta)\right)\leq\beta,
\label{equation: dual3}\end{eqnarray} since $G(u\vert\theta)\cap
J^{-1}(A)\ne\varnothing$ and $J\circ G(u\vert\theta)\cap
A\ne\varnothing$ are equivalent. Finally, by a third application of
P1, (\ref{equation: dual3}) is equivalent to $\beta$ weakly
dominating the minimum of the quantity
$\mathbb{E}(1\{\tilde{Z}\notin J\circ G(\tilde{U}\vert\theta)\})$
over all pairs $(\tilde{Z},\tilde{U})$ with $\tilde{Z}\sim Q$ and
$\tilde{U}\sim\nu(\cdot\vert\theta)$ and the result follows.
\end{proof}

To illustrate the composition theorem, consider a special case of the revealed preference example~\ref{example: revealed preferences} combined with measurement error, as in example~\ref{example: measurement}. Suppose we observe the share
$Y$ of risky assets in the portfolio of investors, who are assumed
to maximize the expectation of a CARA utility function
$u(Y,A;U)=\exp(-U[(1-Y)+YA])$, hence they are assumed to maximize
$Y\mathbb{E}(A)-U Y^2\mbox{var(A)}/2$, where $\mathbb{E}(A)$ is the
perceived mean of the risky asset $A$ and $\mbox{var}(A)$ its
perceived variance. We further suppose investors differ by their
risk aversion $U$, for which the analyst hypothesizes an exponential
distribution ($F_U(u;\theta)=\mathbb{P}(U\leq u;\theta)=1-e^{-\theta
u}$) and by their perception of the riskiness of the asset, and all
the analyst knows is a pair of bounds
$(\underline\lambda,\overline\lambda)$ such that
$\mathbb{E}(A)/\mbox{var}(A)\in[\underline\lambda,\overline\lambda]$.
The investor's maximization yields $Y=\mathbb{E}(A)/U\mbox{var}(A)$,
so that the model can be summarized by $Y\in
G(U)=[\underline\lambda/U,\overline\lambda/U]$. Values
$\underline\lambda=50\%$ and $\overline\lambda=200\%$ can be calibrated
according to \cite{Weitzman:2007}. The true distribution of
income $Y$ is unknown, but the true cumulative distribution of a mismeasured version $Z=Y+\epsilon$, with $\|\epsilon\|\leq\eta$ a.s., is $F_Z(y)=\mathbb{P}(Z\leq z)=\exp(-1/z)$. By Theorem~\ref{theorem: composition},
the identified set $\tilde\Theta_I$ can be derived from the composed correspondence $J\circ G: u\rightrightarrows J\circ G(u)=[\underline\lambda/u-\eta,\overline\lambda/u+\eta]$, where $J: y\rightrightarrows J(y)=B(y,\eta)$ is a dilation satisfying Assumption~\ref{assumption: dilation}.
The
cumulative distribution of risk aversion satisfies $1-e^{-\theta u}=\mathbb{P}(U\leq
u)\in[\mathbb{P}(\overline\lambda/u+\eta\leq
Z),\mathbb{P}(\underline\lambda/u-\eta\leq
Z)]=[1-e^{-(\overline\lambda/u+\eta)^{-1}},1-e^{-(\underline\lambda/u-\eta)^{-1}}]$.
Hence, for all $u>0$, $(\overline\lambda+\eta u)^{-1}\leq\theta\leq(\underline\lambda-\eta u)^{-1}$.
Therefore, the identified set can be derived as
$\tilde \Theta_I=[1/\overline\lambda,1/\underline\lambda]$.

\section{Dilation method and sampling uncertainty}
\label{section: sampling dilation}

\subsection{Confidence regions}\label{subsection: confidence
region} The main application of the Composition
Theorem~\ref{theorem: composition} that we consider here is the
construction of valid confidence regions for partially identified
models, based on a sample of realizations of the observable
variables.

\begin{assumption}[Sampling] \label{assumption: sampling}
Let $(Y_1,\ldots,Y_n)$ be a sample
of independent and identically distributed random vectors with
distribution $P$ and let $P_n=\sum_{j=1}^n\delta_{Y_j}$ be the
empirical distribution associated with the sample.
\end{assumption}

We propose a new method to construct a confidence
region for the identified set $\Theta_I$ of
definition~\ref{definition: identified set}.

\begin{definition}[Confidence region]\label{definition: confidence
region} A valid $\alpha$-confidence region for the identified set
$\Theta_I$ is a sequence of random regions $\Theta_n^\alpha$
satisfying
\[\lim\inf_n\;\mathbb{P}\left(\Theta_I\subseteq\Theta_n^\alpha\right)
\geq1-\alpha.\]\end{definition}

As noted in \cite{IM:2004}, this is not the only way to define
confidence regions in a partially identified setting, as one might
also consider coverage (point wise or uniform) of each value within
the identified set. Here we concentrate on a situation where one
cannot assume that any value within the identified set can be
construed as the true value, so that the whole set is the object of
interest. Moreover, a confidence region for the identified set is
also a uniform confidence region for each of its elements. The
construction of the confidence region is based on a new
nonparametric way of controlling sampling uncertainty and its
validity relies on a corollary to the Composition Theorem
(Theorem~\ref{theorem: composition} of Section~\ref{section:
composition}). We construct sample based sets $J_n^\alpha$, where
$\alpha\in(0,1)$ is the desired confidence level, to account for the
discrepancy between the empirical distribution $P_n$ associated with
the sample and the true observable distribution $P$. We thereby
obtain an analogue of Assumption~\ref{assumption: dilation}:

\begin{assumption}[Sample dilation]
With probability $1-\alpha_n$ such that $\lim\sup_n$
$\alpha_n\leq\alpha$, conditionally on the sample
$(Y_1,\ldots,Y_n)$, the sequence of correspondences $J_n^\alpha$
satisfies $Y\in J_n^\alpha(\tilde{Y}^\ast)$ almost surely for some
$\tilde{Y}\sim P$, $\tilde{Y}^\ast\sim P_n$. \label{assumption:
sample dilation}\end{assumption}

Heuristically, the region $J_n^\alpha$ satisfying
Assumption~\ref{assumption: sample dilation} ensures that with
suitable confidence, the realizations of the empirical distribution
are \emph{caught} by the \emph{enlarged} realizations of the true
distribution $J_n^\alpha(\tilde{Y})$. Once the dilation $J_n^\alpha$
is obtained, the Composition Theorem can be applied to prove the
following:

\begin{theorem} Under assumptions~\ref{assumption: specification},
\ref{assumption: correspondence}, \ref{assumption: parametric},
\ref{assumption: sampling} and~\ref{assumption: sample dilation},
then
$\Theta_n^\alpha:=\{\theta\in\Theta\;\vert\;\;\exists\tilde{Y}^\ast\sim
P_n,\;\tilde{U}\sim\nu(\cdot\vert\theta):\;\mathbb{P}(\tilde{Y}^\ast\notin
J_n^\alpha\circ G(\tilde{U}\vert\theta))=0 \}$ is a valid
$\alpha$-confidence region for the identified set $\Theta_I$.
\label{theorem: dilation}\end{theorem}

The dilation $J_n^\alpha$ is chosen to control the confidence level:
indeed, by Proposition~1 of \cite{GH:2006d} (called P1 in the proof
of Theorem~\ref{theorem: composition}), the statement
$\exists\tilde{Y}^\ast\sim
P_n,\;\tilde{U}\sim\nu(\cdot\vert\theta):\;\mathbb{P}(\tilde{Y}^\ast\notin
J_n^\alpha\circ G(\tilde{U}\vert\theta))=0$ is equivalent to
$P(A)\leq P_n(J_n^\alpha(A)),$ for all Borel subset $A$ of
$\mathcal{Y}$. Hence, the unknown distribution $P$ of an event $A$
is dominated by the empirical distribution of the dilation
$J_n^\alpha(A)$ of the event $A$. As both $P_n$ and
$\nu(\cdot\vert\theta)$ are known, the construction of
$\Theta_n^\alpha$ is feasible and efficient methods to compute it
were proposed in \cite{GH:2008a} and \cite{BMM:2008}.

\subsection{Oracle dilation}\label{subsection: Infeasible
dilation}

We now turn to the question of how to construct the dilation
$J_n^\alpha$ that satisfies Assumption~\ref{assumption: sample
dilation}. When $Y$ is a random variable, such dilation will be
obtained from uniform confidence bands for the quantile process.

\begin{definition}[Quantile process]
Let $F$ be the cumulative distribution of $Y$. Let $Q(t)$,
$t\in[0,1]$ be the quantile function of $Y$, defined by
$Q(t)=\inf\{y\in[0,1]:\,F(y)\geq t\}$. Call $Q_n$ the empirical
quantile relative to the sample $(X_1,\ldots,X_n)$. It is defined by
$Q_n(t)=Y_{(j)}$ for $j-1< nt\leq j$ for each $j$, with $Y_{(j)}$
denoting the $j$th order statistic. The quantile process is defined
as $q_n(t):=\sqrt{n}\left(Q_n(t)-Q(t)\right)$.\end{definition}

The idea of the construction of dilations satisfying
Assumption~\ref{assumption: sample dilation} is based on the
quantile transformation. Indeed, letting $Z$ be a uniform random
variable on $[0,1]$ and defining $\tilde{Y}=Q(Z)$ and
$\tilde{Y}^\ast=Q_n(Z)$, we have a pair of random variables
$\tilde{Y}$ and $\tilde{Y}^\ast$ with respective probability
distributions $P$ and $P_n$. Suppose a uniform confidence band is
available for the quantile function of the form
$\mathbb{P}(\eta_n:=\sup_{0\leq t\leq 1}|q_n(t)|\leq \tilde
c_n(\alpha))=1-\alpha_n$. Then, with probability $1-\alpha_n$, we
have $|\tilde Y^\ast-\tilde Y|=|Q_n(Z)-Q(Z)|\leq \tilde
c_n(\alpha)/\sqrt{n}$ almost surely. Hence, the dilation
$J_n^\alpha$ defined for all $y$ by $J_n^\alpha(y)=B(y,\tilde
c_n(\alpha)/\sqrt{n})$ satisfies Assumption~\ref{assumption: sample
dilation}. Moreover, the choice of dilation
$J_n^\alpha(y)=B(y,\tilde c_n(\alpha)/\sqrt{n})$ is optimal in the
sense that, under the regularity conditions of
Assumption~\ref{assumption: quantile bands}, $|Q_n(Z)-Q(Z)|$
achieves the minimum of $|\tilde Y^\ast-\tilde Y|$ when $\tilde
Y^\ast$ (respectively $\tilde Y$) ranges over the set of random
variables with distribution $P_n$ (respectively $P$). Note that
smaller dilations are desirable, as they maximize informativeness of
the resulting confidence region.

The following conditions guarantee the existence of such uniform
confidence bands for the quantile process.

\begin{assumption}[Uniform quantile bands] The sample
$\{Y_1,\ldots,Y_n\}$ is an iid sample of random variables with
cumulative distribution function $F$ satisfying:
\vskip5pt
\noindent (i) $F(y)$ is twice continuously differentiable on its support $(a,b)$. 
\vskip5pt\noindent (ii) $F^\prime=f>0$ on $(a,b)$. \vskip5pt\noindent (iii) For some $\gamma>0$, $\sup_{y\in(a,b)}F(y)(1-F(y))|f^\prime(y)|/f(y)^2\leq\gamma$. \vskip5pt\noindent (iv) $\lim\sup_{y\downarrow a}f(y)<\infty$ and $\lim\sup_{y\uparrow b}f(y)<\infty$. \vskip5pt\noindent (v) $f$ is nondecreasing (resp. nonincreasing) on an interval to the right of $a$ (resp. to the left of $b$).
\label{assumption: quantile bands}\end{assumption}

A distribution function $F$ satisfying Assumption~\ref{assumption:
quantile bands} is called {\em tail monotonic with index $\gamma$}
by Parzen \cite{Parzen:79}. To indicate the mildness of
Assumption~\ref{assumption: quantile bands}, \cite{Parzen:79} gives
the following example where it fails: $1-F(y)=\exp(-y-C\sin y)$ with
$0.5<C<1$. As shown below, under Assumption~\ref{assumption:
quantile bands}, asymptotic results on the empirical quantile
process allow us to derive a dilation $J_n^\alpha$ that satisfies
Assumption~\ref{assumption: sample dilation} for all
$\alpha\in(0,1)$. Define $c(\alpha)$ implicitly by
$\mathbb{P}(\sup_{0\leq t\leq 1}|B(t)|\leq c(\alpha))=1-\alpha$,
where $B(t)$ is a Gaussian process called a {\em Brownian bridge}.
For any $\alpha\in (0,1)$, we then have the following result.

\begin{proposition}[Oracle dilation]\label{proposition: quantile dilation}
Under assumptions~\ref{assumption: sampling} and~\ref{assumption:
quantile bands}, the dilation $J^\alpha_n$ defined for each $y$ by
$J^\alpha_n(y)=[\;y-c(\alpha)/\sqrt{n}f(y),\;y+c(\alpha)/\sqrt{n}f(y)]$
satisfies Assumption~\ref{assumption: sample
dilation}.\end{proposition}

\begin{proof}[Proof of Proposition \protect\ref{proposition: quantile dilation}] Under Assumption~\ref{assumption: quantile bands}, we have the following strong approximation result in \cite{Csorgo:83}, theorem~4.1.2 page 31: \[
\sup_{0\leq t\leq 1}\vert
f(Q(t))q_n(t)-B_n(t)\vert=O(n^{-1/2+\varepsilon}),\;\;\mbox{a.s.}\]
for $\epsilon>0$ arbitrary, where $B_n(t)$ is a sequence of Brownian
bridges. Hence, the interval \[Q_n(t)-c(\alpha)/\sqrt{n}f(Q(t))\leq
Q(t)\leq Q_n(t)-c(\alpha)/\sqrt{n}f(Q(t))\] is an asymptotically
valid uniform confidence band for $Q(t)$, $0\leq t\leq 1$, of level
$1-\alpha.$ Take $Z$ a uniform random variable on $[0,1]$. Define
$\tilde{Y}^\ast:=Q_n(Z)$ and $Y:=Q(Z)$. By the quantile transform,
$\tilde Y^\ast$ has distribution $P_n$ and $\tilde Y$ has
distribution $P$. Therefore, with probability tending to $1-\alpha$,
there exists $\tilde{Y}^\ast\sim P_n$ and $\tilde Y\sim P$ such that
$\tilde Y-c(\alpha)/\sqrt{n}f(\tilde Y) \leq\tilde Y^\ast\leq \tilde
Y+c(\alpha)/\sqrt{n}f(\tilde Y)$ almost surely, and the result
follows.
\end{proof}

The dilation in proposition~\ref{proposition: quantile dilation}
is infeasible, as it depends on the unknown $f$ and it relies on
quantiles $c(\alpha)$ that are difficult to compute. We develop a
feasible alternative in our dilation bootstrap procedure in
section~\ref{subsection: dilation bootstrap1}. We resort to a
\emph{bootstrap matching algorithm} to construct feasible versions
of the dilation above.

\subsection{Bootstrap dilation}
\label{subsection: dilation bootstrap1}

\begin{figure}[htbp]
\begin{center}
\vskip10pt
\includegraphics[width=12cm]{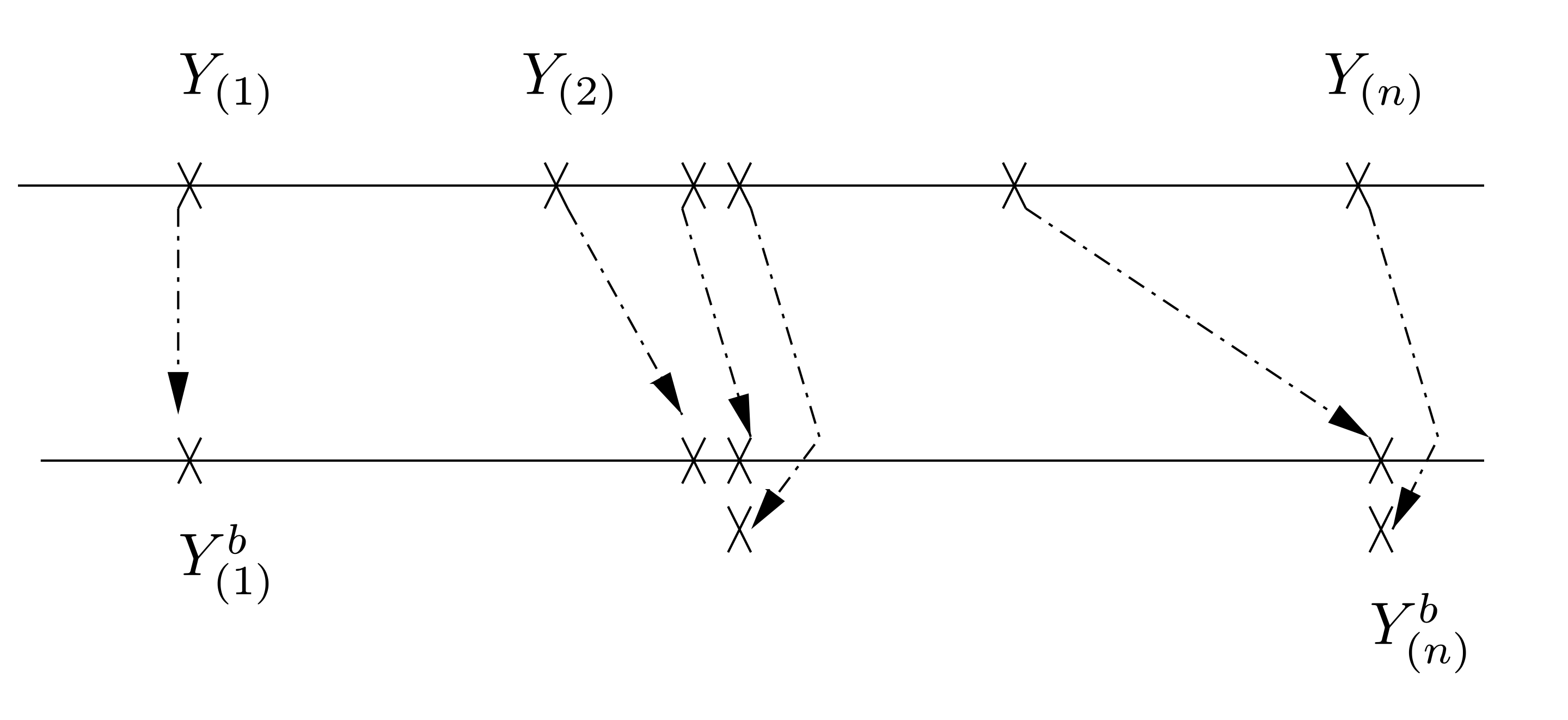} \caption{Bootstrap Quantile Matching.}
\label{figure: bijection}
\end{center}
\end{figure}

To introduce the simple idea underlying the method, consider the
sample $(Y_1,\ldots,Y_n)$ and a given bootstrap realization
$(Y_1^b,\ldots,Y_n^b)$ as in figure~\ref{figure: bijection}. As
before, $(Y_{(1)},\ldots,Y_{(n)})$ are the order statistics
associated with the sample and $(Y_{(1)}^b,\ldots,Y_{(n)}^b)$ are
the order statistics associated with the bootstrap realization (with
arbitrary ranking of the ties). In the illustrative example of
figure~\ref{figure: bijection}, the smallest observation of the
initial sample $Y_{(1)}$ was drawn once in the bootstrap sample, the
second smallest was not drawn, the third smallest was drawn once,
the fourth smallest twice, and the largest $Y_{(n)}$ was drawn
twice. The arrows in the figure represent the bijection that matches
the $j$'th order statistic of the initial sample $Y_{(j)}$ with the
$j$'th order statistic of the bootstrap sample $Y_{(j)}^b$ for each
$j=1,\ldots,n$.

To achieve a bootstrap analog of Assumption~\ref{assumption: sample
dilation}, we need a dilation $J_n^b$ and a permutation $\sigma$ of
$\{1,\ldots,n\}$ such that $Y_{(j)}\in J_n^b(Y_{\sigma(j)}^b)$ for
all $j=1,\ldots,n$. One such permutation matches the order
statistics of the initial sample with the order statistics of the
bootstrap sample. In this matching in the example of
figure~\ref{figure: bijection}, $Y_{(1)}$ is matched with
$Y_{(1)}^b$, namely with itself. $Y_{(2)}$ was not drawn in the
bootstrap sample, so it is matched with $Y_{(2)}^b$, which is equal
to $Y_{(3)}$, for whom $Y_{(2)}$ is the second closest neighbor in
Euclidian distance. $Y_{(3)}$ is the nearest neighbor of its match
$Y_{(3)}^b=Y_{(4)}$, $Y_{(4)}$ is matched with itself, $Y_{(n-1)}$
is the nearest neighbor of its match $Y_{(n-1)}^b=Y_{(n)}$ and
finally $Y_{(n)}$ is matched with itself. The longest distance
between two matches is
$\eta_n^b=|Y^{b}_{(n-1)}-Y_{(n-1)}|=|Y_{(n)}-Y_{(n-1)}|$. Hence, if
$J_n^b(y)=B(y,\eta_n^b)$, $Y^b_{(j)}\in J_n^b(Y_{(j)})$ will be
satisfied for all $j=1,\ldots,n$ in this particular bootstrap sample
realization. The chosen matching in figure~\ref{figure: bijection}
characterizes the bootstrap quantile process (see
Definition~\ref{definition: bootstrap quantile}) and it minimizes
the largest deviation $\eta_n^b$, and hence produces the smallest
dilation.

\begin{definition}[Bootstrap quantile process]
A bootstrap sample is a sample $(Y_1^b,\ldots,Y_n^b)$ of i.i.d.
variables with distribution $P_n$. The quantile function of the
distribution of the bootstrap sample ({\em bootstrap quantile}) is
defined for each $t\in[0,1]$ by $Q_n^b(t)=Y_{(j)}^b$ for $j-1<
nt\leq j$. The {\em bootstrap quantile process} is defined as
$q^b_n(t):=\sqrt{n}(Q_n^b(t)-Q_n(t))$. Call $\eta_n^b$ the maximum
of the bootstrap quantile process. \label{definition: bootstrap
quantile}\end{definition}

In the illustrative example of figure~\ref{figure: bijection}, the
bootstrap quantile process attains its maximum over $t\in[0,1]$ at
$t$ such that $n-2<nt\leq n-1$ and
$\eta_n^b=Y^b_{(n-1)}-Y_{(n-1)}=Y_{(n)}-Y_{(n-1)}$. In the
population of bootstrap realizations, $\eta_n^b$ has distribution
with $1-\alpha$ quantile $c_n^\ast(\alpha)$. The latter can be
approximated by simulation with a large number $B$ of bootstrap
replications. We obtain $\eta_n^b$ for each $b=1,\ldots,B$. Call
$\hat{c}_n^\ast(\alpha)$ the $[B\alpha]$-th largest among the
$\eta_n^b$'s (where $[.]$ denoted integer part) and $\hat
J_n^{\alpha,\ast}(y)=B(y,\hat{c}_n^\ast(\alpha))$, then by
construction, a proportion $1-\alpha$ of the bootstrap samples
indexed by $b=1,\ldots,n$ will satisfy $Y^b_{(j)}\in
J_n^{\alpha,\ast}(Y_{(j)})$ for all $j=1,\ldots,n$. By Theorem~2 of
\cite{Singh:81} (see also Theorem~5.1 of \cite{BF:81}), the
bootstrap quantile process $(q_n^b(t))_{t\in[0,1]}$ has almost
surely the same uniform weak limit as the empirical quantile process
$(q_n(t))_{t\in[0,1]}$ and we therefore have the following result on
the validity of the bootstrap dilation.

\begin{proposition}[Bootstrap dilation]\label{proposition: bootstrap
dilation} Let $c_n^\ast(\alpha)$ be the $1-\alpha$ quantile of the
supremum $\eta_n^b$ of the bootstrap quantile process
$(q_n^b(t))_{t\in[0,1]}$. Under assumptions~\ref{assumption:
sampling} and~\ref{assumption: quantile bands}, the dilation defined
for each $y$ by $J_n^{\alpha,\ast}(y)=B(y,c_n^\ast(\alpha)/\sqrt n)$
satisfies Assumption~\ref{assumption: sample dilation} almost
surely.
\end{proposition}

Note that in the univariate case, the simulation approximation
$\hat{c}_n^\ast(\alpha)$ to the quantile $c_n^\ast(\alpha)$ is very
simple to derive. The simplest algorithm requires ordering the
initial sample and each of the bootstrap samples and computing the
maximum of $|Y^b_{(j)}-Y_{(j)}|$ over $j=1,\ldots,n$. However, we
have introduced, with figure~1 and the discussion above, an
equivalent algorithm, which runs as follows: for each
$b=1,\ldots,B$, find the permutation $\sigma$ over $\{1,\ldots,n\}$,
which minimizes the quantity $\max_j|Y^b_{j}-Y_{\sigma(j)}|$. Unlike
the algorithm based on the order statistics, such an {\em optimal
matching} or {\em optimal assignment} procedure can be performed
regardless of dimension and efficient algorithms and implementations
are available.

\section{Simulation evidence}
\label{section: simulation}

We assess the small sample performance of the dilation bootstrap on the following simulation design. Observable variables $Y$ have a standard normal distribution, while unobserved heterogeneity variable $U$ is assumed to follow a normal distribution with mean $\theta$ and variance $1$. The cumulative distribution of $U$ is denoted $F_U$. The model correspondence $G$ is defined for each $u$ by $G(u)=[u-1,u+1]$, so that the model is characterized by the relation $Y\in G(U)=[U-1,U+1]$.
Therefore, the identified set can be immediately derived as
$\Theta_I=[-1,1]$. For $5,000$
initial samples of size $n=50,100,500$, with empirical
distributions $P_n$, we compute $\hat{c}_n^\ast(\alpha)$ with
$5,000$ bootstrap replications, and use the dilation $\hat
J_n^{\alpha,\ast}(y)=B(y,\hat{c}_n^\ast(\alpha))$, so that a
parameter value $\theta$ belongs to the $(1-\alpha)$-confidence
region $\Theta_{\mathrm{CR}}$ for $\Theta_I$ if and only if there
exist $\tilde{Y}^\ast\sim P_n$ and $\tilde{U}\sim F_U(.;\theta)$
such that $\mathbb{P}^\ast(\tilde{Y}^\ast\in \hat
J_n^{\alpha,\ast}\circ G(\tilde{U}))=1$. Since $P_n$ and $F_U$ are
known, the latter condition can be checked efficiently with the core
determining class method of \cite{GH:2008a}, section~2.3. We report
Monte Carlo coverage probabilities in case of significance level
$\alpha=0.01$, $0.05$ and $0.1$ in Table~\ref{table: dilation}.

\begin{table}
\vskip15pt \caption{Rejection levels from the dilation bootstrap procedure.} \label{table: dilation}\begin{center}
\begin{tabular}{l||c|c|c}
{Sample Size} &50&100&500\\ \hline\hline\\ {$\alpha=0.01$}
&0.0122&0.0118&0.0108\\ \hline\\
{$\alpha=0.05$} &0.0324&0.0364&0.0438 \\ \hline\\ {$\alpha=0.10$} &0.0590&0.0648&0.0754 \\
\end{tabular}
\end{center}
\end{table}

The most notable feature to note is the tendency to under reject in small samples, especially for true size $\alpha=0.10$ but also for true size $\alpha=0.05$. For true size $\alpha=0.01$ on the other hand, the procedure displays slight over rejection in small samples.
For comparison purposes, we also
report coverage probabilities from the generic subsampling procedure
for set coverage in \cite{CHT:2007} based on the criterion function
$\sqrt{n}\max(\max_{j=1,\ldots,n}[F_n(Y_j)+F_U(Y_j+1)],
\max_{j=1,\ldots,n}[-F_n(Y_j)+F_U(Y_j-1)])$. In
order to avoid artificially favoring our results, we report coverage
probabilities under several subsample sizes and when the
true identified set is known and no initial estimate is needed in
the \cite{CHT:2007} procedure. The Monte Carlo coverage probabilities for $500$ initial samples and $500$ subsamples of sizes $40,45,48$ when $n=50$, $85,92,95$ for $n=100$ and $425,450,475$ for $n=500$ are reported in Table~\ref{table: CHT}.
We find the procedure over rejects in all but one case, and there is moderate dependence in the choice of subsample size.

\begin{table}
\vskip15pt \caption{Rejection levels from the infeasible CHT procedure.} \label{table: CHT}\begin{center}
\begin{tabular}{c|c||c|c|c|c|c}
{Sample}& {Subsample}&{$\alpha=0.01$}&{$\alpha=0.05$}&{$\alpha=0.10$}\\ \hline\hline\\\\
&40&0.022&0.086&0.102\\
50&45&0.026&0.060&0.098\\ &48&0.030&0.058&0.130\\ \hline\\
&85&0.040&0.100&0.116\\
100&92&0.034&0.082&0.156\\ &95&0.056&0.080&0.138\\
\hline\\ &425&0.068&0.098&0.140\\ 500&450&0.042&0.084&0.116\\
&475&0.062&0.118&0.138\\
\end{tabular}
\end{center}
\end{table}

\section{Extensions}\label{section: extensions}
The dilation method and dilation bootstrap have natural extensions to the cases, where observable variables $Y$ are multivariate and to the case, where $Y$ is discrete. We consider both extensions in the following subsections.

\subsection{Multivariate extension}\label{section: higher dimensions}
Consider first the case, where the random vector of
observable variables $Y$ has dimension $d\geq2$. This extension
allows the consideration of multiple equations models. Moreover, it
is particularly relevant in this partially identified framework, as
it also allows the consideration of single equations models with
endogenous regressors.

\begin{example}[Single equation model with endogeneity]
Suppose the econometric model under consideration is
$Z=f(X,U;\theta)$, where $Z$ and $X$ are observed random variables,
$U$ is unobserved heterogeneity and $f$ is a function parameterized
by $\theta$. Suppose no assumption is made on the dependence between
$X$ and $U$. Define $Y=(X,Z)^\prime$. Define the correspondence $G$
for each $u$ by $(x,z)\in G(u;\theta)$ if and only if
$z=f(x,u;\theta)$. Then the model can be rewritten $Y\in
G(U;\theta)$ as in Assumption~\ref{assumption: specification}.
\end{example}

In case $Y$ is multivariate, although Theorem~\ref{theorem:
dilation} holds irrespective of dimension, the construction of a
dilation satisfying Assumption~\ref{assumption: sample dilation} can
no longer rely on the traditional quantile process as in
Propositions~\ref{proposition: quantile dilation}
and~\ref{proposition: bootstrap dilation}. However, the quantity
$\eta_n=\inf\{\|\tilde{Y}^\ast-\tilde{Y}\|_\infty:\;\tilde{Y}^\ast\sim
P_n,\;\tilde{Y}\sim P\}$ is still well defined. When attained, it is
achieved by a pair of random vectors $(\tilde{Y}^\ast,\tilde{Y})$
with marginal distributions $P_n$ and $P$, which minimizes the
largest deviation. Equivalently, there exist $\tilde{Y}^\ast\sim
P_n$ and $\tilde{Y}\sim P$ such that $\tilde{Y}^\ast$ belongs to a
closed ball $B(\tilde{Y},\eta_n)$ centered on $\tilde{Y}$ and with
radius $\eta_n$, i.e. such that $\mathbb{E}[1\{\tilde{Y}^\ast\notin
B(\tilde{Y},\eta_n)\}]=0$.

When $Y$ is uniformly distributed on the unit cube $[0,1]^d$, the
quantity $\eta_n$ is well studied in the probability literature.
Hence, using asymptotic results on the quantity $\eta_n$ in the
literature, specifically \cite{LS:89} for the case $d=2$ and
\cite{SY:91} for the case $d\geq3$, we can derive analytical
formulae for the dilation $J_n$:

\begin{proposition}[Minimax matchings]\label{proposition: minimax}
The exist a constant $c>0$ and a function $c_d>0$ of the dimension
$d$ of $Y$ such that $J_n(y)=B(y,c_2(\ln n)^{3/4}/\sqrt{n})$
satisfies Assumption~\ref{assumption: sample dilation} with
$\alpha_n=n^{-c\sqrt{\ln n}}$ when $d=2$ and $J_n(y)=B(y,c_d\d(\ln
n/n)^{1/d})$ satisfies Assumption~\ref{assumption: sample dilation}
for any $\alpha\in[0,1)$ when $d\geq3$.\end{proposition}

However, the results of Proposition~\ref{proposition: minimax} only
pertain to the uniform case and produce conservative confidence
regions. More generally, we propose constructing suitable dilations
based on the distribution of $\eta_n$.

\begin{definition}[Minimax matching]\label{definition: minimax}
Call $c_n(\alpha)$ the $1-\alpha$ quantile of the distribution of
$\eta_n=\inf\{\|\tilde{Y}^\ast-\tilde{Y}\|_\infty:\;\tilde{Y}^\ast\sim
P_n,\;\tilde{Y}\sim P\}$.\end{definition}

By construction, we then see that the ball $B(y,c_n(\alpha))$ is a
suitable dilation, in the sense that it satisfies
Assumption~\ref{assumption: sample dilation}.

\begin{proposition}[Multivariate oracle dilation]
\label{proposition: multivariate dilation} The dilation $J^\alpha_n$
defined for each $y$ by $J^\alpha_n(y)=B(y,c_n(\alpha))$ satisfies
Assumption~\ref{assumption: sample dilation}.\end{proposition}

As for the approximation of $c_n(\alpha)$ to obtain a feasible
dilation, once again, although the quantile process is no longer
defined, the matching algorithm described in
Section~\ref{subsection: dilation bootstrap1} is easily
generalizable and delivers a bootstrap dilation approximation of
$J_n^\alpha$. The general procedure is described as follows.
\vskip10pt\begin{center}{\sc Bootstrap Algorithm:}\end{center}
\begin{itemize} \item Consider bootstrap samples $(Y^b_{1},\ldots,Y^b_{n})$,
$b=1,\ldots,B$ drawn from $P_n$ and call $P_{n}^b$ the empirical
distribution of sample $b$. \item For each bootstrap replication
$b$, define
\[\eta_n^b=\min_\sigma\max_{j\in\{1,\ldots,n\}}\|Y^b_j-Y_{\sigma(j)}\|,\]
where $\sigma$ ranges over all permutations of $\{1,\ldots,n\}$.
\item Let $\hat{c}_n^\ast(\alpha)$ be the $[B\alpha]$ largest among the
$\eta_n^b$, $b=1,\ldots,B$, and for each $y$, set $\hat
J_n^{\alpha,\ast}(y)=B(y,\hat{c}_n^\ast)$. \end{itemize} The problem
of finding the permutation that achieves $\eta_n^b$ is called {\em
bottleneck bipartite matching} in the combinatorial optimization and
operations research literature.

\subsection{Case of discrete choice}
We now turn to the case of aggregate data from discrete choice. To
fix ideas, consider a voting model, where $K$ parties are
represented in $n$ electoral districts and observations $\hat
p_{i,k}$, $i=1,\ldots,n$ and $k=1,\ldots,K$, are reported shares of
votes for party $k$ in district $i$. Voter $l$ chooses the party that
maximizes their utility
$u^l_{i,k}(\theta)+\rho_{i,k}+\epsilon^l_{i,k}$, where
$u_{i,k}(\theta)$ is a deterministic  function of (observed
covariates and) the unknown parameter $\theta$, $\rho_{i,k}$ are
random district-party effects (independent of voters) and the
$\epsilon^l_{i,k}$'s are i.i.d. type I extreme value random
utilities. True vote shares for party $k$ in district $i$ satisfy
$\ln p_{i,k}^\ast(\rho_{i,k})=u_{i,k}(\theta)+\rho_{i,k}+
\ln\sum_k\exp(u_{i,k}+\rho_{i,k})$. True shares $p_{i,k}^\ast$ are
unobserved, however, due to the possibility of electoral fraud.
Reported shares $p_{i,k}$ are assumed to satisfy $p_{i,k}\geq
p_{i,k}^\ast$ when a representative of party $k$ is present during
the vote count in district $i$. In districts, where no party representative is present, the situation is equivalent to missing data on vote shares. Let $X_{i,k}$ be equal to $1$ if a
representative of party $k$ is present in district $i$ during vote
count, and zero otherwise. We assume
$X=(X_{i,k})_{i=1,\ldots,n;\;k=1,\ldots,K}$ is exogenous. The
correspondence characterizing the model is
\begin{eqnarray*}G\left((\rho_{i,k})_{k=1}^{K}\vert
X;\theta\right)=\hskip200pt\\\left\{(p_{i,k})_{k=1}^{K}:\;
\sum_{k=1}^{K}p_{i,k}=1;\; p_{i,k}\geq p_{i,k}^\ast(\rho_{i,k})
X_{i,k},\mbox{ each }k\right\}.\end{eqnarray*} District $i$ has
$n_i$ voters. Call $\hat p_{i,k}$ the proportion of votes in
district $i$ reported as going to party $k$ and write $\hat
p_i=(\hat p_{i,k})_{k=1,\ldots,K}$. By the central limit theorem,
$\sqrt{n_i}(\hat p_i-p_i)$ has Gaussian limiting distribution with
zero mean and covariance matrix $V_i$, with diagonal elements
$p_{i,k}(1-p_{i,k})$ and off-diagonal elements $-p_{i,k}p_{i,k'}$.
Call $Z_i$ a random vector with distribution $N(0,V_i/n_i)$ and let
$\eta_i$ be such that $\mathbb{P}(Z_i\notin B(0,\eta_i))=\alpha_i$,
where $B(0,\eta_i)$ is the open ball centered at zero with radius
$\eta_i$. Define the dilation $J_{n_i}^{\alpha_i}$ defined for each
$p$ by $J_{n_i}^{\alpha_i}(p)=B(p,\eta_i)$. Then
$J_n^\alpha(p)=\bigcup_iJ_{n_i}^{\alpha_i}(p)=B(p,\max_i\eta_i)$
satisfies Assumption~\ref{assumption: sample dilation} for
$\alpha=\lim\sup_n\Pi_{i=1}^n\alpha_i$.

In the two-party case, call $\hat p_i$ the reported share of votes
for party~1, $p_i$ the {\em true} or {\em population} reported share
and $p_i^\ast$ the true share (absent reporting fraud). The true
share satisfies $\ln
p_i^\ast(\rho_i)=u_{i,1}(\theta)+\rho_{i,1}+\ln(\exp(u_{i,1}(\theta)+
\rho_{i,1})+ \exp(u_{i,2}(\theta)+\rho_{i,2}))$. Because of fraud
issues, all we know about the relation between $p_i$ and $p_i^\ast$
is the following:
\begin{eqnarray*} p_i\geq p_i^\ast&&\mbox{ if party~1 places an
observer in district }i.\\p_i\leq p_i^\ast&&\mbox{ if party~2 places
an observer in district }i.\end{eqnarray*} Note that reported vote shares
are equal to true vote shares in case both parties have observers present for vote count. Letting $X_{i,k}$ take
value $1$ if party $k$ places an observer in district $i$ and zero
otherwise, the correspondence characterizing the model is
\begin{eqnarray*}G(\rho_i\vert X,\theta)&=&\left\{p_i:\; p_i\geq
p_i^\ast(\rho_i)X_{i,1}\mbox{ and
}(1-p_i)\geq(1-p_i^\ast(\rho_i))X_{i,2}\right\}\\&=&
\left[\frac{X_{i,1}\exp(u_{i,1}(\theta)+\rho_{i,1})}{\exp(u_{i,1}(\theta)+
\rho_{i,1})+
\exp(u_{i,2}(\theta)+\rho_{i,2})},\right.\\&&\hskip30pt\left.
1-\frac{X_{i,2}\exp(u_{i,1}(\theta)+ \rho_{i,2})}{
\exp(u_{i,1}(\theta)+\rho_{i,1})+\exp(u_{i,2}(\theta)+\rho_{i,2})}\right].
\end{eqnarray*}
By the central limit theorem, $\sqrt{n_i}(\hat p_i-p_i)$ has
Gaussian limiting distribution with zero mean and variance
$p_i(1-p_i)$. Call $c_{\alpha_i/2}$ the quantile of level
$1-\alpha_i/2$ of the standard normal distribution. Call
$\eta=\max_i\eta_i$ with
$\eta_i=c_{\alpha_i/2}\sqrt{p_i(1-p_i)/n_i}$. Then the dilation
defined for each $p$ by $J_n^\alpha(p)=[p-\eta,p+\eta]$ satisfies
Assumption~\ref{assumption: sample dilation} with
$\alpha=\lim\sup_n\Pi_i\alpha_i$. The composition of the dilation
$J_n^\alpha$ and the correspondence $G$ yields \[J_n^\alpha\circ
G(\rho\vert
X;\theta)=[X_1p^\ast(\rho)-\eta,X_2(1-p^\ast(\rho))+\eta].\] The
region $\tilde \Theta_I$ containing all $\theta$ such that $\hat
p\in J_n^\alpha\circ G(\rho\vert X,\theta)$ a.s. is therefore a
valid confidence region for the identified set and can be computed
efficiently using methods proposed in \cite{GH:2008a}.

\section*{Conclusion}
We have proposed a method to combine several sources of uncertainty,
such as missing or corrupted data and structural incompleteness in
the model through a composition of correspondences. We show that our
composition theorem applies in particular to the construction of
confidence regions in partially identified models of general form.
In that case, the composition theorem is applied to the composition
of the correspondence that defines the econometric structure and a
dilation of the sample space that controls the significance level
and allows to replace the unknown distribution of observable data by
the empirical distribution of the sample in the characterization of
compatibility between model and data. An important computational
advantage of this method over previous proposed confidence regions
for partially identified parameters is that the dilation is
performed independently of the structural parameter, hence needs to
be performed only once. The remaining search over the parameter
space is purely deterministic. The dilation is obtained through a
minimax matching procedure. It is equivalent to a uniform confidence
band for the quantile process when the dimension of the endogenous
variable is one, however, it has no parallel in higher dimensions.
The method is shown to perform well in simulation experiments.

\section*{Acknowledgements}
We thank Christian Bontemps, Gary Chamberlain, Victor Chernozhukov,
Pierre-Andr\'e Chiappori, Ivar Ekeland, Rustam Ibragimov, Guido
Imbens, Thierry Magnac, Francesca Molinari, Alexei Onatski, Geert
Ridder, Bernard Salani\'e, participants at the ``Semiparametric and
Nonparametric Methods in Econometrics'' conference in Oberwolfach
and seminar participants at BU, Brown, CalTech, Chicago, \'Ecole
polytechnique, Harvard, MIT Sloan, Northwestern, Toulouse, UCLA,
UCSD and Yale for helpful comments (with the usual disclaimer). Both
authors gratefully acknowledge financial support from NSF grant SES
0532398 and from Chaire AXA ``Assurance des Risques Majeurs'' and
Chaire Soci\'et\'e G\'en\'erale ``Risques Financiers''. Galichon's
research is partly supported by Chaire EDF-Calyon ``Finance and
D\'{e}veloppement Durable'' and FiME, Laboratoire de Finance des
March\'es de l'Energie. Henry's research is also partly supported by SSHRC Grant 410-2010-242.

\printbibliography

\end{document}